\documentclass[conference]{IEEEtran}
\IEEEoverridecommandlockouts
\usepackage{cite}
\usepackage{amsmath,amssymb,amsfonts}
\usepackage{algorithmic}
\usepackage{graphicx}
\usepackage{textcomp}
\usepackage{xcolor}
\usepackage{tikz}
\usepackage{booktabs}
\usepackage{enumitem}
\usetikzlibrary{positioning,arrows.meta,shapes.geometric,calc}
\def\BibTeX{{\rm B\kern-.05em{\sc i\kern-.025em b}\kern-.08em
    T\kern-.1667em\lower.7ex\hbox{E}\kern-.125emX}}

\begin{document}

\title{Pluralistic Human-Robot Interaction: Designing for Robot Interaction with Diverse Communities}

\author{
\IEEEauthorblockN{Raj Korpan}
\IEEEauthorblockA{\textit{Hunter College \& The Graduate Center, City University of New York}\\
raj.korpan@hunter.cuny.edu}
}

\maketitle

\begin{abstract}
Social robots are being developed for homes, schools, and other environments where they will interact with diverse users. While Human-Robot Interaction (HRI) research often emphasizes natural communication, engagement, personalization, and task success, these goals do not fully address the social complexity of real-world deployment. This paper proposes \emph{Pluralistic HRI}, a framework for designing social robots that treat human diversity as a foundational design concern. The framework brings together pluralism, civic dialogue, perspective-taking, empathy, intercultural competence, cultural humility, and moral imagination to guide inclusive, adaptive, and ethically grounded interaction. We outline how pluralistic HRI can inform design, evaluation, and deployment in diverse human communities.
\end{abstract}

\begin{figure*}
\centering
\includegraphics[width=0.96\linewidth]{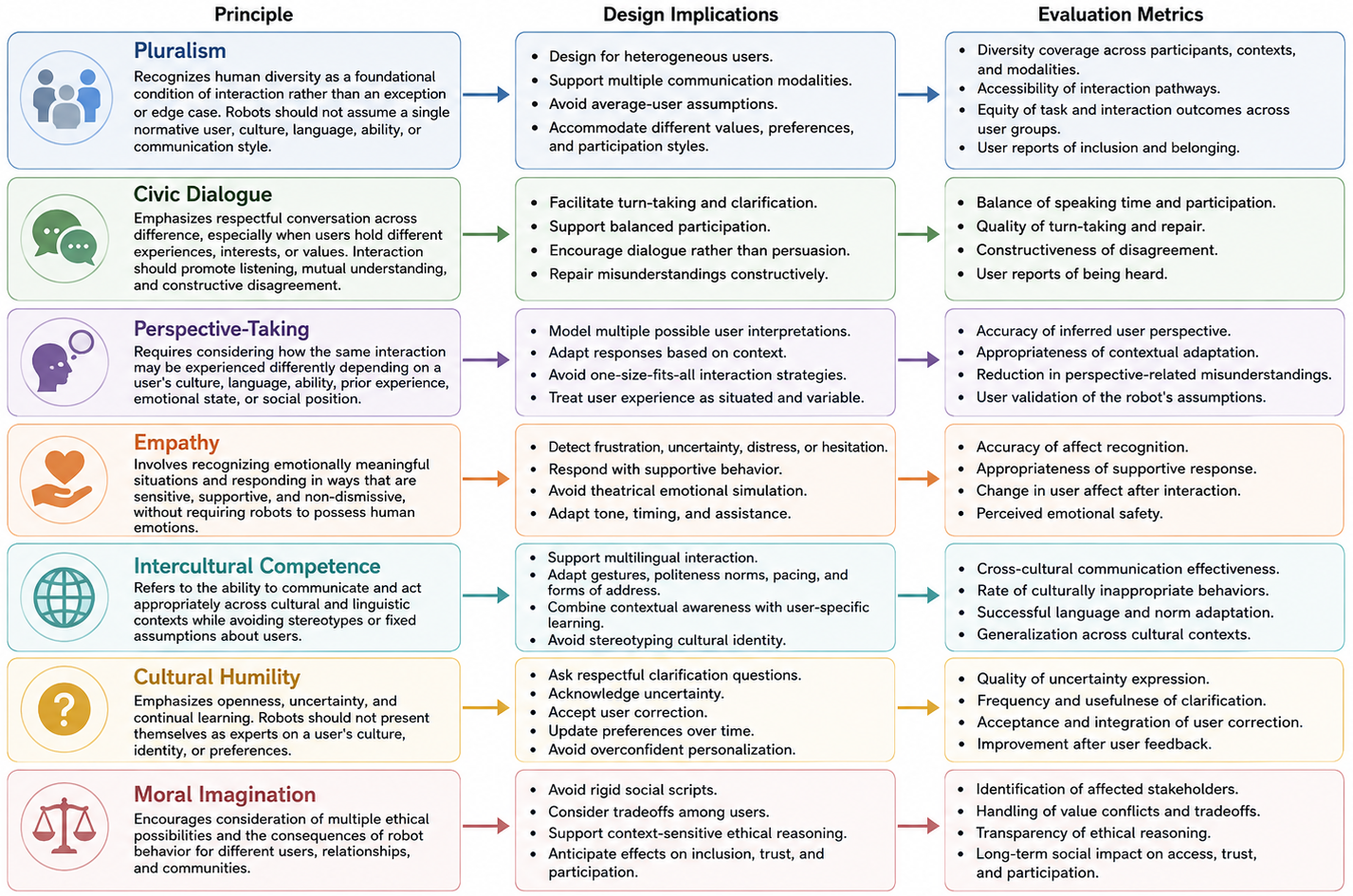}
\caption{Pluralistic HRI principles with corresponding design implications and evaluation metrics.}
\label{fig:pluralistic-hri-principles}
\end{figure*}

\section{Introduction}

Social robots are being developed for everyday environments~\cite{zawieska2023towards,hornecker2022beyond}. In these settings, robots will encounter people who differ in language, culture, age, disability, communication style, social norms, emotional needs, and lived experience~\cite{seaborn2023not,lim2021social,nam2026not}. This diversity is not an edge case for social robotics; it is the standard condition of real-world deployment~\cite{seaborn2023not,hornecker2022beyond,tanevska2023inclusive}.
Much HRI research emphasizes natural communication, intent recognition, engagement, personalization, and task success. While important, these goals do not fully capture the social complexity of public and community-facing environments~\cite{serholt2022introduction,ljungblad2024critical,ostrowski2022ethics}. A robot may be conversationally fluent while assuming a narrow model of the user, efficient while failing to recognize exclusion, personalized while reinforcing stereotypes, or clear in communication while avoiding disagreement, uncertainty, and moral complexity~\cite{winkle2023feminist,zhu2024robots,coggins2023seven,williams2023voice}.

This paper argues that social robots require a broader design paradigm: \emph{Pluralistic Human--Robot Interaction} (HRI), an approach to robot communication and interaction that treats human diversity as a foundational design concern and supports respectful, inclusive, and ethically aware engagement across difference. The framework brings together values and capacities often discussed separately, including civic dialogue, perspective-taking, empathy, intercultural competence, cultural humility, and moral imagination. These ideas shift HRI from communication as information exchange toward interaction as relational participation in diverse human communities.

\section{Pluralistic HRI as a Unifying Framework}

Pluralism refers to the coexistence of multiple identities, values, cultures, beliefs, abilities, and ways of life within a shared society~\cite{eck2006pluralism,soutphommasane2005grounding}. In HRI, this means robots should not be designed around a presumed average user or a single normative model of social behavior~\cite{seaborn2023not,winkle2023feminist,zhu2024robots,korpan2023trust}. They should instead be designed for environments in which people may communicate differently, interpret behavior differently, disagree about values, and require different forms of support~\cite{lim2021social,hornecker2022beyond,coggins2023seven,nam2026not}.
Current social robots often rely on assumptions about what counts as natural, polite, helpful, or appropriate interaction~\cite{hornecker2022beyond,serholt2022introduction}. These assumptions may work in narrow settings but become fragile across diverse communities. Eye contact, turn-taking, directness, emotional expression, personal space, and forms of address can vary across cultures, abilities, age groups, and individual preferences~\cite{lim2021social,seaborn2023not,ornelas2023redefining}. A pluralistic approach therefore treats interaction norms as situated rather than universal~\cite{vsabanovic2010robots,larsson2023towards}.

Pluralistic HRI provides a framework for designing robots that can participate constructively in diverse human communities. As shown in Figure~\ref{fig:pluralistic-hri-principles}, the framework organizes seven principles: pluralism, civic dialogue, perspective-taking, empathy, intercultural competence, cultural humility, and moral imagination. These principles are mutually reinforcing dimensions of socially responsible interaction. Pluralism defines the overall orientation, while the remaining principles describe the capacities needed to enact it in practice. Civic dialogue supports respectful conversation across difference~\cite{mccoy2002deliberative}; perspective-taking helps robots consider how the same interaction may be experienced differently~\cite{batson1997perspective,trafton2005enabling}; empathy supports sensitive responses to emotionally meaningful situations~\cite{cuff2016empathy,paiva2017empathy,park2022empathy}; intercultural competence guides adaptation across language and cultural norms~\cite{leung2014intercultural,bruno2019knowledge}; cultural humility encourages uncertainty, clarification, and user correction~\cite{foronda2016cultural}; and moral imagination helps designers anticipate the broader consequences of robot behavior~\cite{friedman2019value,johnson2014moral}.

These principles also translate into concrete design and evaluation concerns. Pluralistic HRI shifts design away from one-size-fits-all interaction and toward systems that ask, clarify, adapt, support multiple forms of participation, and remain responsive to feedback. It also expands evaluation beyond task completion and user satisfaction to include inclusion, dialogue quality, contextual appropriateness, emotional safety, intercultural effectiveness, responsiveness to correction, and ethical handling of tradeoffs.

\section{Discussion}

Pluralistic HRI reframes diversity not as a demographic variable to be checked after system development, but as a condition that should shape interaction models from the beginning. This requires robots to move beyond average-user assumptions and toward systems that can ask clarifying questions, accommodate multiple communication modalities, adapt to situated norms, and accept user correction~\cite{seaborn2023not,winkle2023feminist,zhu2024robots,lim2021social}. It also requires evaluation practices that look beyond task success, conversational naturalness, and aggregate satisfaction~\cite{ostrowski2022ethics,ostrowski2022design,serholt2022introduction}. A pluralistic approach examines the equity of outcomes across users, the extent to which people feel heard and respected, the appropriateness of the robot’s responses to emotional and cultural context, and the system’s ability to handle disagreement, uncertainty, and value conflict without defaulting to rigid or exclusionary behavior.

Future work should develop interaction models that represent diverse user perspectives, communication norms, and accessibility needs, along with datasets that better reflect multilingual, multicultural, intergenerational, queer, and disability-inclusive interaction~\cite{seaborn2023not,saettone2026diversity,bruno2023culture,nam2026not,korpan2025encoding}. In this sense, pluralistic HRI provides a bridge between technical work on dialogue, multimodal interaction, and personalization and broader commitments to accessibility, inclusion, and socially responsible design~\cite{friedman2013value,winkle2023feminist,zhu2024robots,tanevska2023inclusive}.

This framework positions social robots not merely as tools or conversational agents, but as participants in shared social environments whose behavior can shape trust, access, identity, participation, and belonging over time~\cite{vsabanovic2010robots,hornecker2022beyond,jarske2025could}. As robots enter community spaces, their interaction patterns may influence who feels comfortable using them, whose needs are recognized, and whose ways of communicating are treated as legitimate~\cite{lim2021social,ostrowski2022ethics,zhu2024robots}. For example, a humanoid robot in a community health clinic could enact pluralistic HRI by combining speech, gesture, gaze, facial expression, and on-screen text to ask how a user prefers to receive information rather than assuming a single communication norm. If the user avoids eye contact, requests another language, pauses before responding, or corrects the robot’s form of address, the robot could adapt its behavior and treat these cues as invitations for clarification rather than as interaction failures.

Pluralistic HRI encourages researchers and designers to consider both the risks and possibilities of robot behavior in diverse communities. The risks include reinforcing stereotypes, excluding non-normative users, and privileging dominant cultural assumptions~\cite{seaborn2023not,winkle2023feminist,zhu2024robots,korpan2025encoding}. The possibilities include robots that support civic dialogue, assist diverse users in accessing services, mediate participation in group settings, and adapt respectfully across linguistic, cultural, emotional, and accessibility differences~\cite{jarske2025could,lim2021social,bruno2019knowledge}. By treating pluralism as a core requirement for social robots rather than as a specialized accessibility or ethics add-on, HRI can move beyond the goal of natural conversation toward systems that are more inclusive, adaptive, reflective, and ethically grounded.

\bibliographystyle{ieeetr}
\bibliography{mypapers,bushi}

@inproceedings{winkle2023feminist,
  title={Feminist human-robot interaction: Disentangling power, principles and practice for better, more ethical HRI},
  author={Winkle, Katie and McMillan, Donald and Arnelid, Maria and Harrison, Katherine and Balaam, Madeline and Johnson, Ericka and Leite, Iolanda},
  booktitle={Proceedings of the 2023 ACM/IEEE international conference on human-robot interaction},
  pages={72--82},
  year={2023}
}

@inproceedings{zhu2024robots,
  title={Robots for social justice (r4sj): Toward a more equitable practice of human-robot interaction},
  author={Zhu, Yifei and Wen, Ruchen and Williams, Tom},
  booktitle={Proceedings of the 2024 ACM/IEEE International Conference on Human-Robot Interaction},
  pages={850--859},
  year={2024}
}

@article{larsson2023towards,
  title={Towards a socio-legal robotics: a theoretical framework on norms and adaptive technologies},
  author={Larsson, Stefan and Liinason, Mia and Tanqueray, Laetitia and Castellano, Ginevra},
  journal={International Journal of Social Robotics},
  volume={15},
  number={11},
  pages={1755--1768},
  year={2023},
  publisher={Springer Netherlands Dordrecht}
}

@inproceedings{ostrowski2022design,
  title={Design justice for robot design and policy making},
  author={Ostrowski, Anastasia K and Breazeal, Cynthia},
  booktitle={2022 17th ACM/IEEE international conference on human-robot interaction (HRI)},
  pages={1170--1172},
  year={2022},
  organization={IEEE}
}

@article{vsabanovic2010robots,
  title={Robots in society, society in robots: Mutual shaping of society and technology as a framework for social robot design},
  author={{\v{S}}abanovi{\'c}, Selma},
  journal={International Journal of Social Robotics},
  volume={2},
  number={4},
  pages={439--450},
  year={2010},
  publisher={Springer}
}

@inproceedings{ostrowski2022ethics,
  title={Ethics, equity, \& justice in human-robot interaction: A review and future directions},
  author={Ostrowski, Anastasia K and Walker, Raechel and Das, Madhurima and Yang, Maria and Breazea, Cynthia and Park, Hae Won and Verma, Aditi},
  booktitle={2022 31st IEEE international conference on robot and human interactive communication (RO-MAN)},
  pages={969--976},
  year={2022},
  organization={IEEE}
}

@incollection{friedman2013value,
  title={Value sensitive design and information systems},
  author={Friedman, Batya and Kahn Jr, Peter H and Borning, Alan and Huldtgren, Alina},
  booktitle={Early engagement and new technologies: Opening up the laboratory},
  pages={55--95},
  year={2013},
  publisher={Springer}
}

@book{friedman2019value,
  title={Value sensitive design: Shaping technology with moral imagination},
  author={Friedman, Batya and Hendry, David G},
  year={2019},
  publisher={Mit Press}
}

@incollection{bruno2023culture,
  title={Culture in social robots for education},
  author={Bruno, Barbara and Amirova, Aida and Sandygulova, Anara and Lugrin, Birgit and Johal, Wafa},
  booktitle={Cultural robotics: social robots and their emergent cultural ecologies},
  pages={127--145},
  year={2023},
  publisher={Springer}
}

@article{lim2021social,
  title={Social robots on a global stage: establishing a role for culture during human--robot interaction},
  author={Lim, Velvetina and Rooksby, Maki and Cross, Emily S},
  journal={International Journal of Social Robotics},
  volume={13},
  number={6},
  pages={1307--1333},
  year={2021},
  publisher={Springer}
}

@article{jarske2025could,
  title={How could social robots support societal participation? findings from five design workshops with young people},
  author={Jarske, Salla and Kaipainen, Kirsikka and Ahtinen, Aino and Varsaluoma, Jari and V{\"a}{\"a}n{\"a}nen, Kaisa},
  journal={International Journal of Social Robotics},
  volume={17},
  number={4},
  pages={563--585},
  year={2025},
  publisher={Springer}
}

@article{ornelas2023redefining,
  title={Redefining culture in cultural robotics},
  author={Ornelas, Mark L and Smith, Gary B and Mansouri, Masoumeh},
  journal={AI \& SOCIETY},
  volume={38},
  number={2},
  pages={777--788},
  year={2023},
  publisher={Springer}
}

@article{serholt2022introduction,
  title={Introduction: special issue—critical robotics research},
  author={Serholt, Sofia and Ljungblad, Sara and N{\'\i} Bhroin, Niamh},
  journal={AI \& SOCIETY},
  volume={37},
  number={2},
  pages={417--423},
  year={2022},
  publisher={Springer}
}

@inproceedings{zawieska2023towards,
  title={Towards HRI of everyday life: human lived experiences with social robots},
  author={Zawieska, Karolina and Sorenson, Jessica},
  booktitle={Companion of the 2023 ACM/IEEE International Conference on Human-Robot Interaction},
  pages={347--350},
  year={2023}
}

@incollection{ljungblad2024critical,
  title={Critical perspectives in human--robot interaction design},
  author={Ljungblad, Sara and Gamboa, Mafalda},
  booktitle={Designing Interactions with Robots},
  pages={148--160},
  year={2024},
  publisher={Chapman and Hall/CRC}
}

@article{hornecker2022beyond,
  title={Beyond dyadic HRI: building robots for society},
  author={Hornecker, Eva and Krummheuer, Antonia and Bischof, Andreas and Rehm, Matthias},
  journal={interactions},
  volume={29},
  number={3},
  pages={48--53},
  year={2022},
  publisher={ACM New York, NY, USA}
}

@article{soutphommasane2005grounding,
  title={Grounding multicultural citizenship: From minority rights to civic pluralism},
  author={Soutphommasane, Tim},
  journal={Journal of Intercultural Studies},
  volume={26},
  number={4},
  pages={401--416},
  year={2005},
  publisher={Taylor \& Francis}
}

@article{eck2006pluralism,
  title={What is pluralism},
  author={Eck, Diana L},
  journal={Pluralism. org. Available online: http://pluralism. org/pluralism/what\_is\_pluralism (accessed on 17 January 2016)},
  year={2006}
}

@article{coggins2023seven,
  title={The seven troubles with norm-compliant robots},
  author={Coggins, Tom N and Steinert, Steffen},
  journal={Ethics and Information Technology},
  volume={25},
  number={2},
  pages={29},
  year={2023},
  publisher={Springer}
}

@article{mccoy2002deliberative,
  title={Deliberative dialogue to expand civic engagement: what kind of talk does democracy need?},
  author={McCoy, Martha L and Scully, Patrick L},
  journal={National civic review},
  volume={91},
  number={2},
  pages={117--135},
  year={2002},
  publisher={Wiley Subscription Services, Inc., A Wiley Company Hoboken}
}

@article{batson1997perspective,
  title={Perspective taking: Imagining how another feels versus imaging how you would feel},
  author={Batson, C Daniel and Early, Shannon and Salvarani, Giovanni},
  journal={Personality and social psychology bulletin},
  volume={23},
  number={7},
  pages={751--758},
  year={1997},
  publisher={Sage Publications Sage CA: Thousand Oaks, CA}
}

@article{trafton2005enabling,
  title={Enabling effective human-robot interaction using perspective-taking in robots},
  author={Trafton, J Gregory and Cassimatis, Nicholas L and Bugajska, Magdalena D and Brock, Derek P and Mintz, Farilee E and Schultz, Alan C},
  journal={IEEE Transactions on Systems, Man, and Cybernetics-Part A: Systems and Humans},
  volume={35},
  number={4},
  pages={460--470},
  year={2005},
  publisher={IEEE}
}

@article{cuff2016empathy,
  title={Empathy: A review of the concept},
  author={Cuff, Benjamin MP and Brown, Sarah J and Taylor, Laura and Howat, Douglas J},
  journal={Emotion review},
  volume={8},
  number={2},
  pages={144--153},
  year={2016},
  publisher={Sage Publications Sage UK: London, England}
}

@article{park2022empathy,
  title={Empathy in human--robot interaction: Designing for social robots},
  author={Park, Sung and Whang, Mincheol},
  journal={International journal of environmental research and public health},
  volume={19},
  number={3},
  pages={1889},
  year={2022},
  publisher={MDPI}
}

@article{paiva2017empathy,
  title={Empathy in virtual agents and robots: A survey},
  author={Paiva, Ana and Leite, Iolanda and Boukricha, Hana and Wachsmuth, Ipke},
  journal={ACM Transactions on Interactive Intelligent Systems (TiiS)},
  volume={7},
  number={3},
  pages={1--40},
  year={2017},
  publisher={ACM New York, NY, USA}
}

@article{leung2014intercultural,
  title={Intercultural competence},
  author={Leung, Kwok and Ang, Soon and Tan, Mei Ling},
  journal={Annu. Rev. Organ. Psychol. Organ. Behav.},
  volume={1},
  number={1},
  pages={489--519},
  year={2014},
  publisher={Annual Reviews}
}

@article{saettone2026diversity,
  title={Diversity and Culture in Social Robotics: A Scoping Review},
  author={Saettone, Lorenza and Sgorbissa, Antonio and Recchiuto, Carmine Tommaso},
  journal={International Journal of Social Robotics},
  volume={18},
  number={3},
  pages={51},
  year={2026},
  publisher={Springer}
}

@article{bruno2019knowledge,
  title={Knowledge representation for culturally competent personal robots: requirements, design principles, implementation, and assessment},
  author={Bruno, Barbara and Recchiuto, Carmine Tommaso and Papadopoulos, Irena and Saffiotti, Alessandro and Koulouglioti, Christina and Menicatti, Roberto and Mastrogiovanni, Fulvio and Zaccaria, Renato and Sgorbissa, Antonio},
  journal={International Journal of Social Robotics},
  volume={11},
  number={3},
  pages={515--538},
  year={2019},
  publisher={Springer}
}

@article{foronda2016cultural,
  title={Cultural humility: A concept analysis},
  author={Foronda, Cynthia and Baptiste, Diana-Lyn and Reinholdt, Maren M and Ousman, Kevin},
  journal={Journal of Transcultural Nursing},
  volume={27},
  number={3},
  pages={210--217},
  year={2016},
  publisher={Sage Publications Sage CA: Los Angeles, CA}
}

@article{seaborn2023not,
  title={Not only WEIRD but “uncanny”? A systematic review of diversity in human--robot interaction research},
  author={Seaborn, Katie and Barbareschi, Giulia and Chandra, Shruti},
  journal={International Journal of Social Robotics},
  volume={15},
  number={11},
  pages={1841--1870},
  year={2023},
  publisher={Springer}
}

@book{johnson2014moral,
  title={Moral imagination: Implications of cognitive science for ethics},
  author={Johnson, Mark},
  year={2014},
  publisher={University of Chicago Press}
}

@inproceedings{korpan2023trust,
  title={Trust in Queer Human-Robot Interaction},
  author={Korpan, Raj},
  booktitle={Trust, Acceptance and Social Cues in Human-Robot Interaction (SCRITA) Workshop Proceedings at RO-MAN 2023},
  year={2023}
}
\end{document}